\documentclass[11pt,twocolumn]{article}
\usepackage{amsmath,amsfonts,amsthm,amssymb,color}
\usepackage{fancybox}
\usepackage{graphicx}
\usepackage{ltexpprt}
\usepackage{subfig}
\usepackage{multirow}
\usepackage[shortlabels]{enumitem}
\setlist[enumerate, 1]{1.,topsep=3pt}
\usepackage[font={small}]{caption}
\newtheorem{theorem}{Theorem}[section]

\newtheorem{lemma}[theorem]{Lemma}

\theoremstyle{definition}

\newenvironment{fminipage}%
  {\begin{Sbox}\begin{minipage}}%
  {\end{minipage}\end{Sbox}\fbox{\TheSbox}}

\newcommand\bb{\boldsymbol{\mathit{b}}}

\newcommand\ff{\boldsymbol{\mathit{f}}}

\newcommand\rr{\boldsymbol{\mathit{r}}}

\newcommand\ww{\boldsymbol{\mathit{w}}}

\newcommand\xx{\boldsymbol{\mathit{x}}}

\newcommand\LL{\boldsymbol{\mathit{L}}}

\newcommand{\figsize}{.36}
\newcommand{\imgsize}{.39}

\begin{document}

\title{An Empirical Study of Cycle Toggling
Based Laplacian Solvers\footnotemark[1]}

\author{Kevin Deweese\footnotemark[2] \\ UCSB \\ kdeweese@cs.ucsb.edu \and
John R. Gilbert \\ UCSB \\ gilbert@cs.ucsb.edu \and
Gary Miller \\ CMU \\ glmiller@cs.cmu.edu \and
Richard Peng \\Georgia Tech \\ rpeng@cc.gatech.edu \and
Hao Ran Xu \\MIT \\ haoranxu510@gmail.com \and
Shen Chen Xu \\ CMU \\ shenchex@cs.cmu.edu
}

\maketitle

\footnotetext[1]{Partially supported by NSF Grants CCF-1637523, CCF-1637564, and  CCF-1637566 titled: AitF: Collaborative Research: High Performance Linear
System Solvers with Focus on Graph Laplacians}
\footnotetext[2]{Partially supported by Intel Corporation}

\begin{abstract}
We study the performance of linear solvers for graph Laplacians
based on the combinatorial cycle adjustment methodology
proposed by [Kelner-Orecchia-Sidford-Zhu STOC-13].
The approach finds a dual flow solution to this linear system
through a sequence of flow adjustments along cycles.
We study both data structure oriented and recursive methods for handling
these adjustments.

The primary difficulty faced by this approach, updating and querying
long cycles, motivated us to study an important
special case: instances where all cycles are formed by fundamental
cycles on a length $n$ path.
Our methods demonstrate  significant speedups over previous
implementations, and are competitive with standard numerical routines.
\end{abstract}

\begin{figure}[htb!]
\centering
\includegraphics[scale=\figsize]{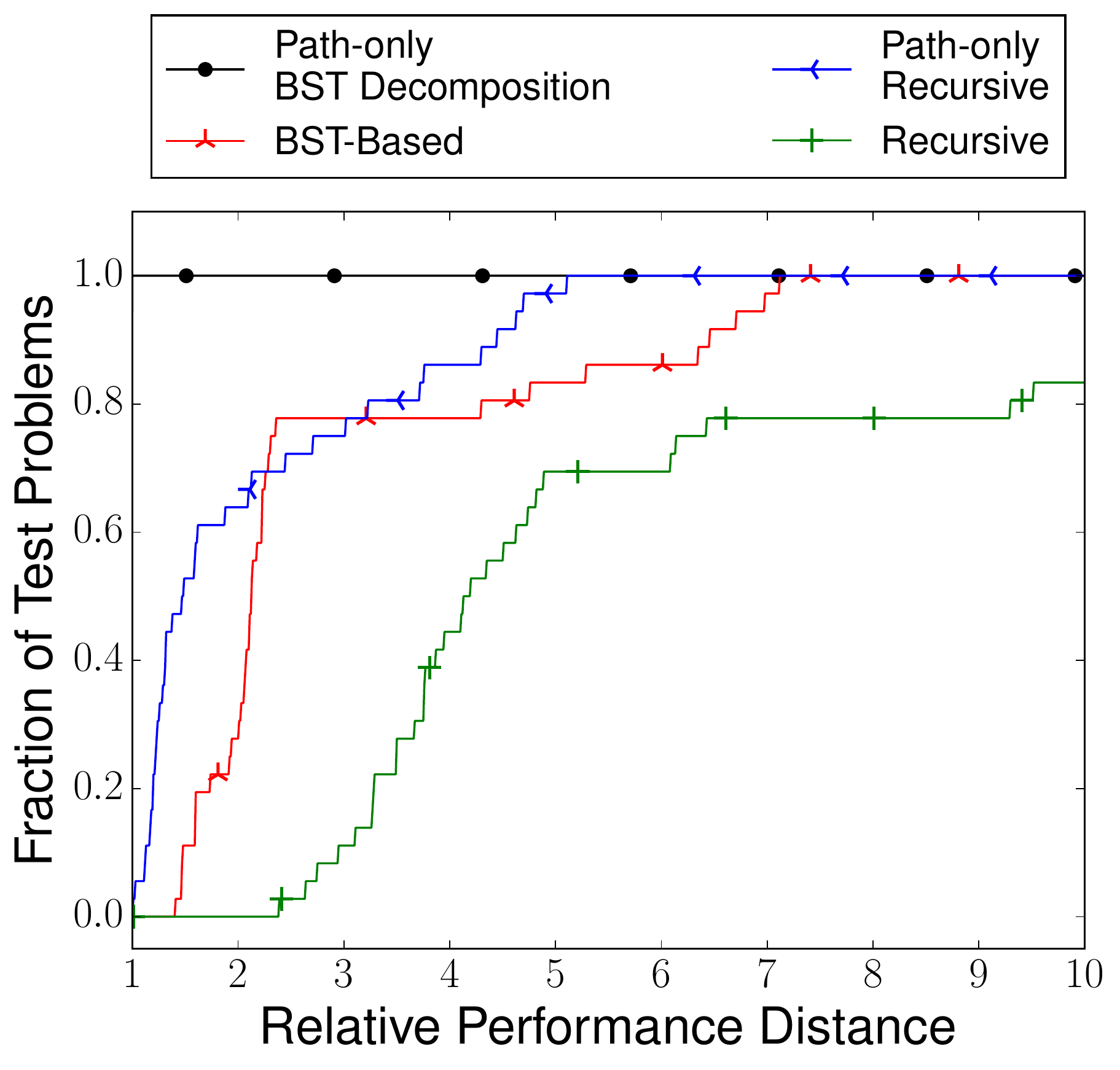}

\caption{Performance profile of cycle-toggle time. The relative performance ratio of a method is its
cycle-toggle time $/$ best cycle toggle time for a single problem.
This plot shows the fraction of problems that are within a distance from this relative performance ratio. The faster a method converges to 1 on this plot, the better its performance relative to the others.}
\label{fig:perfprofile}
\end{figure}

\section{Introduction}
\label{sec:intro}

Much progress has been made recently toward
the development of graph Laplacian linear solvers
that run in linear times polylogarithmic time~\cite{KoutisMP10:journal,
KoutisMP11,KelnerOSZ13,LeeS13,PengS14,CohenKMPPRX14,KyngLPSS15:arxiv}.
These methods use a combination of combinatorial, randomized,
and numerical methods to obtain algorithms that provably
solve {\em{any}} graph Laplacian linear system in
time faster than sorting to constant precision.

Linear solvers for graph Laplacians have
a wide range of applications.
They can be used to solve problems
such as image denoising, finding maximum flows in a graph,
and more generally solving linear programs
with an underlying graph, such as, minimum cost maximum flow and graph theoretic
regression problems~
\cite{BomanHV04,TolliverMiller06,ChristianoKMST11,ChinMMP13,LeeRS13,Madry13,LeeS14,CoFMNPW14,KyngRS15}.
Many of these applications stem from the following connection through
optimization: solving linear systems is equivalent to minimizing
$\ell_2$ norms over a suitable set.
Many applications can in turn be viewed as solving a problem based
on a different norm, such as $\ell_1$ or $\ell_{\infty}$.
The gap between these norms can be addressed through iterative
schemes, leading to algorithms that repeatedly call linear system solvers.

Laplacian linear solvers can be divided into primal solvers, which
solve for a set of vertex potentials, and dual solvers,
which solve for a set of edge flows that minimize energy.
The theoretically fastest known solvers are primal solvers
which use recursive preconditioned
Chebyshev iterations~\cite{CohenKMPPRX14}.
On the other hand, the near-linear time algorithm
with the simplest description works in the dual space~\cite{KelnerOSZ13}.
We believe that the fastest solver will
be one that combines both a potential and flow based approach.
The goal of this paper is to empirically better understand flow based
methods in order to facilitate their integration into primal-dual
algorithmic schemes.

Our main contribution is an experimental investigation of different
cycle-toggling implementations and an examination of the
resulting performance implications. To that end we introduce a class
of synthetic, weighted graphs that are both simple enough to reason
about theoretically, and rich enough to yield interesting behavior
for cycle-toggling implementations.
One of the implementations we use is
a novel divide-and-conquer technique which we describe. 
We end with a comparison of cycle-toggling implementations
to conjugate gradient.

\section{Background}
\label{sec:background}
\subsection{Definitions}
Symmetric diagonally dominant (SDD) matrices and M matrices can be reduced
to Laplacian matrices asymptotically quickly, so the fastest SDD solvers rely on
Laplacian solvers. Laplacians are equivalent to graphs,
which we define as $G = (V, E, \ww)$ where $V$ is a vertex
set, $E$ a set of edges, and $\ww$ a set of edge weights.
The Laplacian is given by
\[
\LL_{i, j} = 
\begin{cases}
\deg(i) & \text{if } i = j\\
-\ww_{ij} & \text{otherwise}
\end{cases},
\]
where $\deg(i)$ is the weighted degree, or sum of incident
edge weights on vertex $i$.
The problem of interest is to solve $L\xx=\bb$ for $\xx$ given
$\bb$.

There is a useful electrical network interpretation of
\[
\xx^T \LL \xx = \sum_{uv \in E}  \ww_{uv} \left( \xx_u - \xx_v \right)^2
\]
where $\xx_u - \xx_v$ can be viewed as voltages~\cite{DoyleS84:book},
and $\ww_{uv}$ represents the inverse of resistance in terms
of energy dissipation.
This definition of resistances gives a corresponding 
electrical flow
interpretation, which forms the basis of the Kelner et
al.s algorithm~\cite{KelnerOSZ13}, which we will call KOSZ.
In this flow interpretation the problem translates to finding a flow $\ff$
that meets demands given by $\bb$, and minimizes
$\sum_{e} \rr_e \ff_e^2$.

\subsection{Existing Methods}
The underlying algebraic operations of theoretically fast graph Laplacian
solvers can be viewed as
either directly manipulating the potential vectors, or the dual flows.
To date, empirical studies of these solvers have focused on the dual flow
based algorithms, leading to mixed results~\cite{BomanDG15,HoskeLMW15,BomanDG16},
most of which are not directly competitive with numerical
methods such as conjugate gradient (CG)~\cite{Saad03}
or multigrid~\cite{BriggsM00}, and instead bound iteration count.
In this paper, we study these dual algorithms with additional insights
obtained during the study of vector based primal algorithms.
We show that the dual adjustment stages can be unraveled in ways
similar to recursive steps in vector solvers.
This allows us to both improve the dual adjustment routine,
as well as having it interact with classical iterative methods
such as conjugate gradient.
Our main experimental results are on improving the performance of
both data structural and recursive approaches, and comparing
their performance to conjugate gradient.

Crucial to the performance of these dual flow solvers is the cycle adjustment
process: here most of the cycles are long, thus cost-prohibitive
to adjust in nearly-linear time.
To obtain nearly-linear performance, these updates are restricted to
fundamental cycles of a tree.
This restriction allows updates to be processed using
tree data structures.
These structures are based on ``virtual tree'' representations of trees
that allow each path to be broken down into $O(\log{n})$ subtrees.
Updating cycles is then done by accessing and modifying the
corresponding labels.
Handling these updates efficiently has proven to be directly related
to the performance of implementations of this algorithm~\cite{KelnerOSZ13}.

\subsection{KOSZ overview}

The KOSZ algorithm~\cite{KelnerOSZ13} randomly selects cycles
and adjusts the flow along a cycle to bring it
to the minimum energy state, while maintaining a feasible flow.
These cycles are formed by first picking a spanning tree $T$.
Then each off-tree edge forms a cycle, known as a
fundamental cycle with respect to $T$. Collectively these
cycles form a fundamental cycle basis which spans the cycle
space of the graph.

Given a cycle of length $k$ with flows $\ff_1 \ldots \ff_k$
oriented in the forward direction, our goal is to find a change
in flow $\Delta$ that minimizes the updated flow energy
\begin{multline*}
\sum_{i} \rr_i \left( \ff_i + \Delta \right)^2
\\= \left( \sum_{i} \rr_i \right) \Delta^2 + 2 \left( \sum_{i} \rr_i \ff_i  \right) \Delta_i + \sum_{i} \rr_i \ff_i^2.
\end{multline*}
This is minimized by setting
\[
\Delta = - \frac{\sum_{i} \rr_i \ff_i}{\sum_{i} \rr_i}.
\]

The choice of cycles to update is dictated by the stretch
of the off-tree edges.
Conceptually, the stretch of an edge is the length of
the detour that must be traversed in the tree if the edge is removed.
This removal ``stretches'' the edge across the new path. 
Here length is measured in terms of resistances, or
inverse edge weight $\frac{1}{\ww_e}$.
For an edge $e$, let the (unique) simple path between
its end points in tree be $P_T(e)$, then
\[
str(e) = \ww_e \sum_{e' \in P_T(e)} \frac{1}{\ww_{e'}}.
\]
The main result of~\cite{KelnerOSZ13} is:
\begin{theorem}
Given a tree with total stretch $S$, repeatedly sampling the edges
randomly with probability proportional to $1 + str(e)$
and bringing the corresponding cycle to its minimum energy
state gives an $1 + \epsilon$-approximate answer in
$O((m + S) \log(1 / \epsilon))$ iterations.
\end{theorem}

\section{Implementing Cycle Toggling}
\label{sec:algorithms}

Cycle-toggling methods require many cycle updates for energy
minimization, necessitating quick update operations.
We need to support the following operations on a tree $T$, where
each edge $e$ is associated with a fixed resistance $\rr_e$ and a flow $\ff_e$:
\begin{enumerate}[itemsep=-1ex]
\item Query: Compute sums of $\rr_e$ and $\rr_e \ff_e$ along a path in $T$.
\item Update: Increment all the flows on a path in $T$ by $\Delta$.
\end{enumerate}

Although these updates are not adaptive,
the result of each update does depend on all
previous updates that interact with the path.
This creates fundamental restrictions on cycle-toggling speed.
This is especially true when considering
any possible parallelism of updating multiple cycles
simultaneously.

In the rest of this section we consider two different
schemes for achieving fast cycle updates.
The first uses data structures similar to the ones used
by the KOSZ algorithm to update each
cycles in $O(\log(n))$ time.
The second is a divide-and-conquer approach we
introduce, which contracts the path based on
preselected cycle updates.

\subsection{Reduction to Balanced BSTs}
We hope to provide the reader with a brief overview of our data structure
approach along with the approach used
by KOSZ~\cite{KelnerOSZ13}.
The KOSZ data structures are based on top-down partitions of trees.
Our implementations are based on a variant of this that uses binary search trees as
building blocks.
To help explain this, we first consider the easier case in which $T$ is just a path,
where we can solve the problem by building a static
balanced binary search tree (BST)~\cite{CormenLRS09}.
Any subtree in the BST corresponds to an interval in the path, 
which can be decomposed into a disjoint union of at most $2\log n$ subtrees 
and nodes in the BST.
To support our query and update operations, we add two pieces of information at every node $v$:
\begin{enumerate}[itemsep=-1ex]
\item
The \textit{sum}, $\sum_{i}r_if_i$ where $i \in$ the subtree containing $v$
\item
A lazy tag $t$, denoting the pending changes of flow in this subtree, caused
by updates to parents.
\end{enumerate}
The BST can answer the interval queries by adding up the \textit{sum} fields of the corresponding subtrees.
Note that this requires the \textit{lazy tag} fields of
all ancestors of the nodes added to be $0$.
This can be handled by `pushing down' such fields as we
access the BST.
The updates involve  modifying the \textit{lazy tag} and \textit{sum} fields of the subtrees 
correspondingly.
This gives us a $O(\log n)$ per operation algorithm for the case where $T$ is a path.
\vspace*{-.3cm}
\begin{figure}[ht]
\centering

\subfloat[Heavy path re-rooted at separator vertex]{\includegraphics[scale=\imgsize]{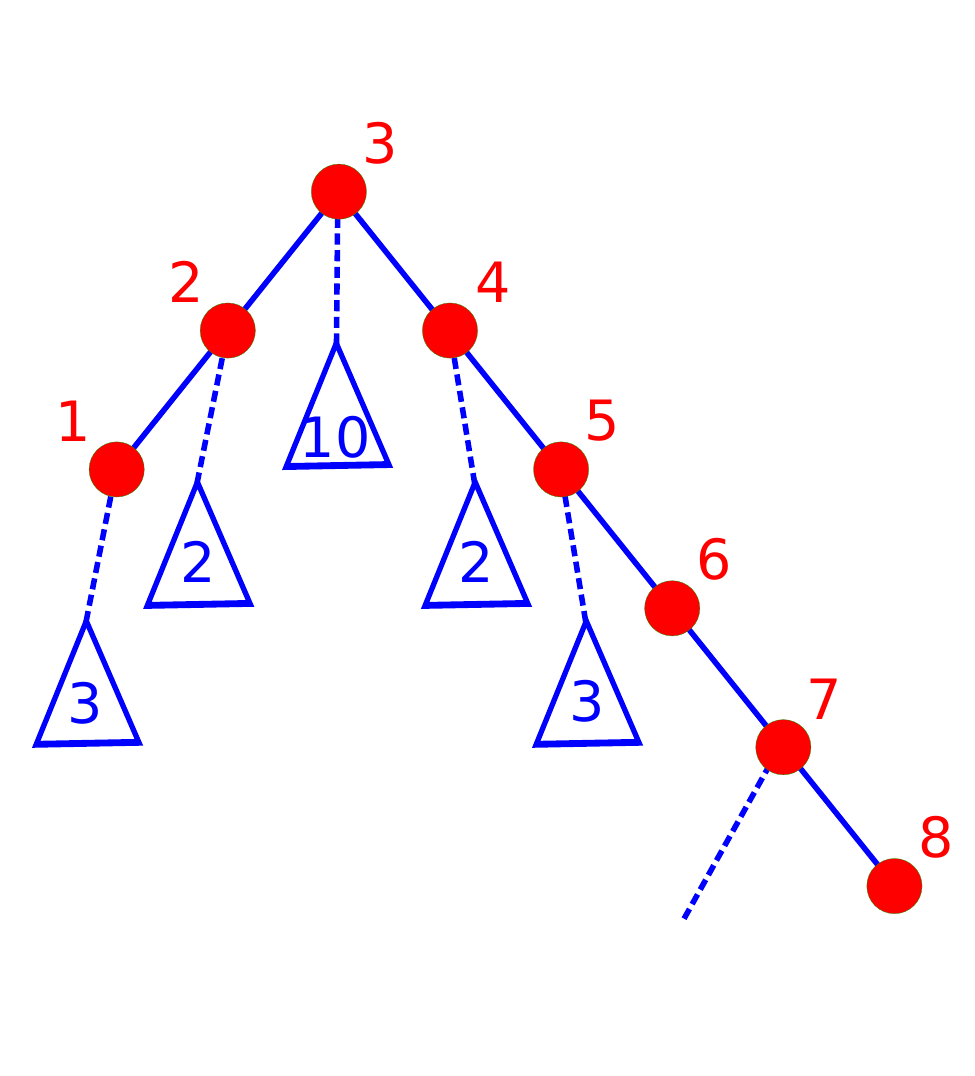}}
\hspace{.6cm}
\subfloat[Virtual tree where heavy path is represented
by the BST that stores it]{\includegraphics[scale=\imgsize]{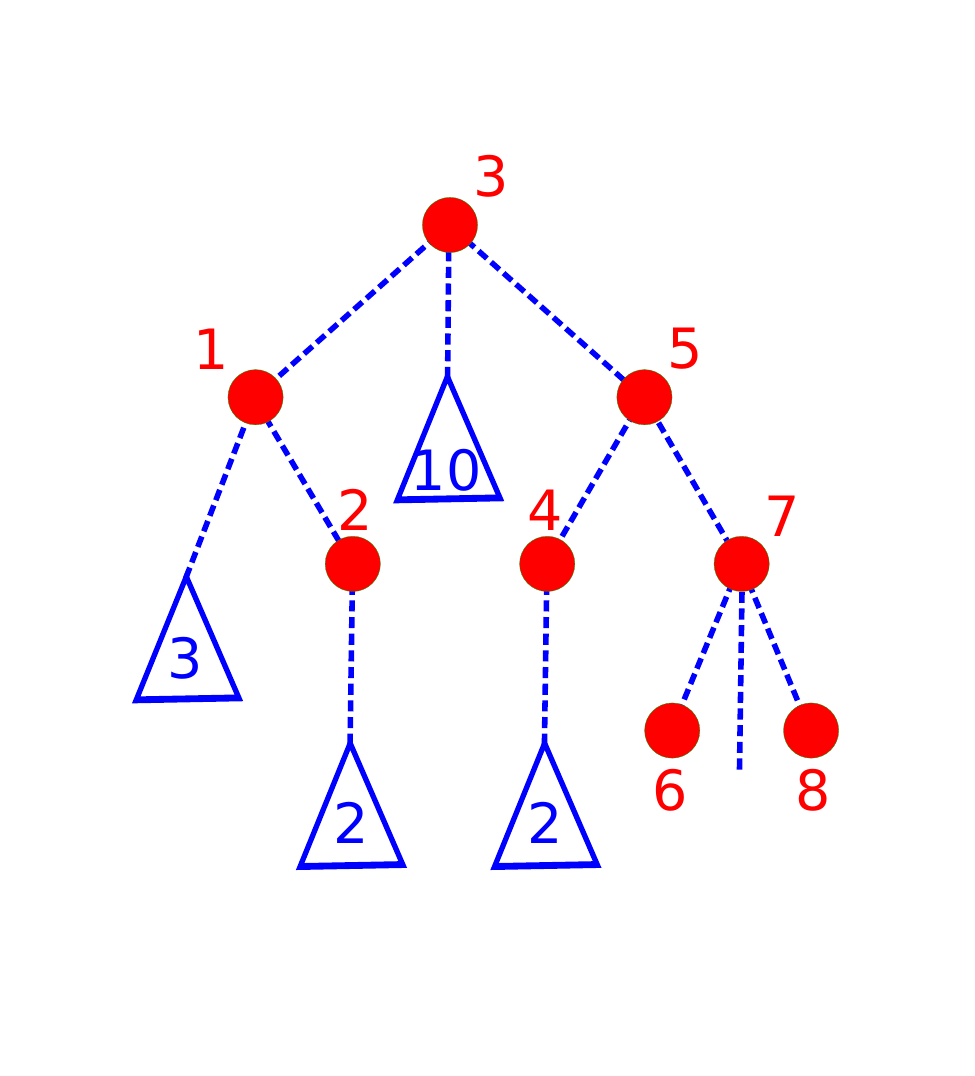}}
\caption{One step of a heavy-light decomposition. Triangles are subtrees labeled with size.}
\label{fig:treedecomp}
\end{figure}

A classic way to generalize the path case to a tree is to use
a heavy-light decomposition (HLD)~\cite{SleatorT83}.
Here, one first arbitrarily roots the tree.
Then for every vertex $u$, 
we denote $v$ as the child of $u$ whose subtree has the
largest size (i.e. contains most vertices).
We mark every edge $(u,v)$ as \textit{heavy} and say that all edges not marked \textit{heavy} 
are \textit{light}. 
An unextendable path of heavy edges is called a \textit{heavy chain}. 
This decomposes the tree into heavy chains and light edges. 

The key fact about this decomposition is that for any vertex $v$, its path
to the root intersects at most $O(\log n)$ heavy chains and $O(\log n)$ light edges.
Therefore, to support query and update operations on a tree,
it suffices to handle the light edges and
support these operations on heavy paths.
For the latter, this is exactly the special path case
and we can use BSTs described above.
This leads to a theoretical time bound $O(\log^2 n)$
per operation, but a quite good running time experimentally.

This method is connected to the data structures used in KOSZ via virtual trees.
Such a tree contains all the BST edges for heavy chains along with light edges.
An example of creating a virtual tree from a HLD is shown in Figure~\ref{fig:treedecomp}.
We can further optimize cycle updates by reducing the virtual tree height.
A path between $u$ and $v$ in the original tree 
can be decomposed into the disjoint union of left-subtrees of nodes in the path
between $u$ and $v$ in the virtual tree.
In HLD, this virtual tree has height $O(\log^2 n)$ (since each BST has height $O(\log n)$ 
and there are at most $O(\log n)$ heavy chains encountered in any path),
so the time bound is $O(\log^2 n)$.

A better virtual tree can be constructed in a recursive manner. 
Consider the heavy chain starting from the root of $T$. 
Using the properties of heavy chains, one can prove that there
exists a node $v$ in the heavy chain, 
whose removal splits $T$ into subtrees which have size at most half of the original tree size. 
We use $v$ as the root of the BST for this heavy chain, and construct recursively.
The virtual tree satisfies the property that any child has at most half the size of its parent, 
so it has height at most $\log n$. 
This gives us a $O(\log n)$ per operation algorithm.
Compared to the recursive-separator based
routine from~\cite{KelnerOSZ13}, this scheme
fixes the heavy path in addition to the root of
the virtual tree.
While this only changes the constants in the analysis,
in terms of implementation it allows us to directly
use the binary tree routine for paths mentioned above.

\subsection{Recursive Divide-and-Conquer}

The other main approach that we explore is a
recursive divide-and-conquer scheme.
The KOSZ solver treats cycle updates as an online
process, a cycle is sampled, then updated, before
another cycle is sampled.
We consider the potential of an offline approach
where we preselect $N$ cycles, and use knowledge
of this set to speed up the update of the set
as a whole.
This method recursively divides the $N$ cycles in half until
the subsets are each of size less than $n$.
The cycles in the last level of the recursion are then updated
in their preselected order.

\begin{figure}[ht]
\centering
\subfloat[Original Graph]{\includegraphics[scale=\imgsize]{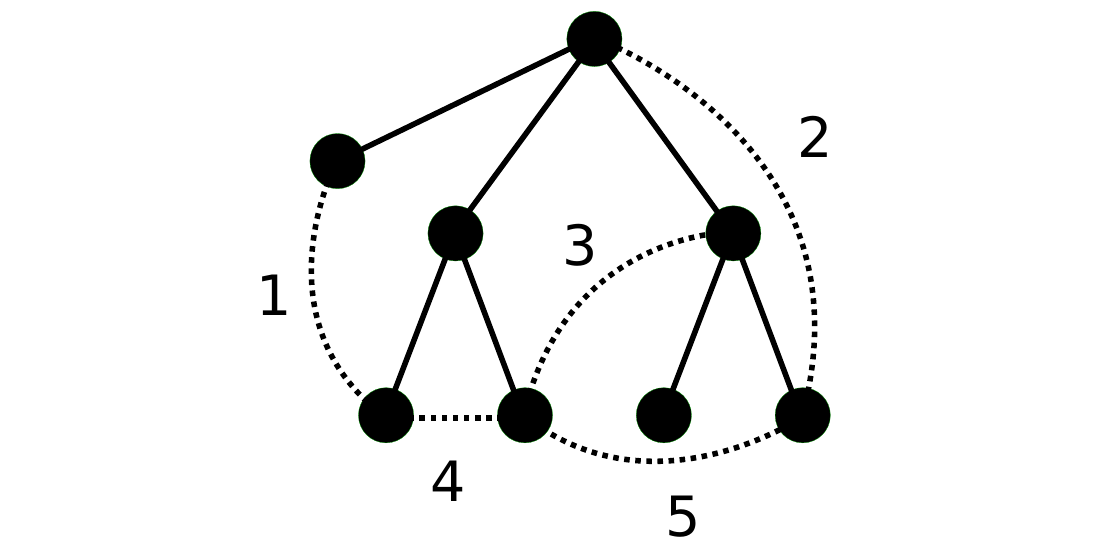}}
\newline
\captionsetup[subfigure]{oneside,margin={-1cm,0cm}}
\subfloat[Subgraph 1,4]{\includegraphics[scale=\imgsize]{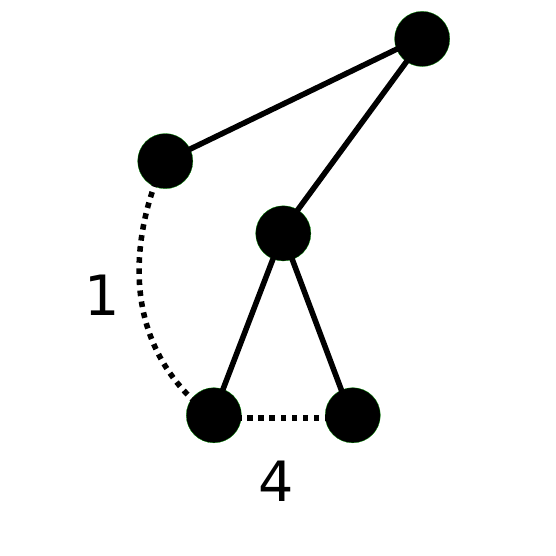}}
\hspace{.7cm}
\captionsetup[subfigure]{oneside,margin={-.4cm,0cm}}
\subfloat[Subgraph 2,3,5]{\includegraphics[scale=\imgsize]{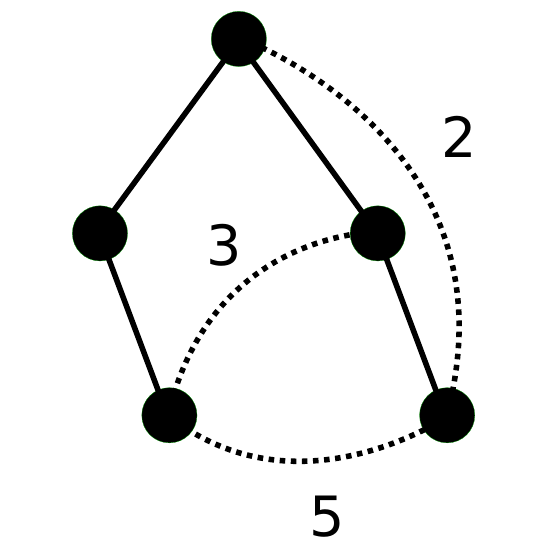}}
\newline
\captionsetup[subfigure]{oneside,margin={-1.4cm,0cm}}
\hspace*{-1.2cm}
\subfloat[Contraction of(b)]{\includegraphics[scale=\imgsize]{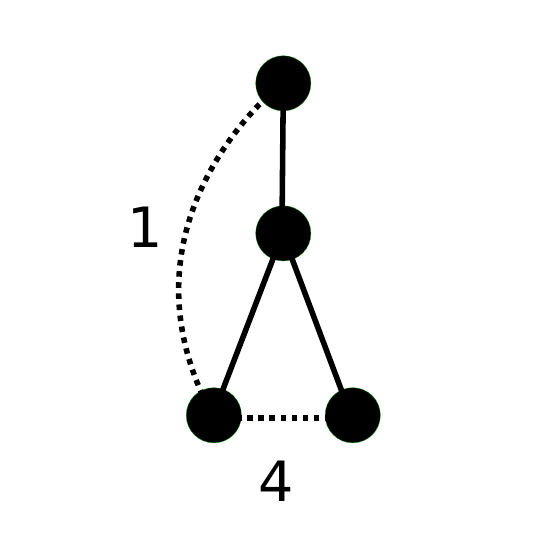}}
\hspace{.8cm}
\captionsetup[subfigure]{oneside,margin={-.7cm,0cm}}
\subfloat[Contraction of(c)]{\includegraphics[scale=\imgsize]{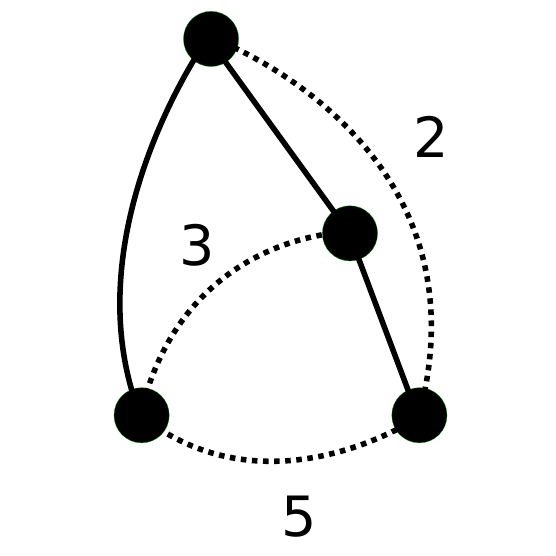}}
\caption{Illustration of graph reduction and contraction in divide-and-conquer.
5 cycles are preselected in the original graph(a) and divided into
two groups, cycles (1,4) and (2,3,5). These cycles induce subgraphs (b,c)
which only include edges and vertices of the relevant cycles.
These subgraphs are then path contracted (d,e) to further reduce size.}
\label{fig:batch}
\end{figure}

The speedup of this approach lies in the fact that we can reduce
the problem to only the part of the graph involved in our preselected updates.
We can further reduce the size of the graph by path contraction,
condensing two edges if they are only updated when the other is updated.
An example of this reduction and contraction is shown in Figure~\ref{fig:batch}.
This process results in several smaller graphs, where the cycles are updated,
before pushing the cycle update information back up the recursive subgraph hierarchy.
As this process resembles the recursive subgraph hierarchy of multigrid methods,
we borrow the terms restriction and prolongation to describe the transfer
of flow information up and down the hierarchy.
This process is more formally captured in the following lemma.

\begin{lemma}
\label{lem:contract}
Given a tree on $n$ vertices, and $N$ cycle updates,
we can form a tree on $3N$ vertices, perform the
corresponding cycle updates on them, and transfer
the state back to the original graph.
Furthermore, both the reduction and prolongation
steps take $O(n)$ time.
\end{lemma}
This procedure is identical to the greedy elimination,
or partial Cholesky factorization steps from the
ultra-sparsification routines~\cite{SpielmanTengSolver:journal}.
Recursively dividing the cycle set yields a recurrence of the form:
\[
\mathcal{T}(N) = O(N) + 2 \mathcal{T}(N /2),
\]
which solves to $\mathcal{T}(N) = O(N \log{N})$.
If we set the size of our preselected cycle set
to $O(n)$, then updating the entire set takes
$O(n \log{n})$ work, leading to a cost of $O(\log{n})$ per update.

Unfortunately, the divide-and-conquer scheme
does not parallelize naturally: the second recursive
call still depends on the outcome of the first one.
Furthermore, the bottleneck of this routine's performance is
the restriction and prolongation steps, which unlike multigrid
can not be reused when we resample another set.
A large part of the expense is that vertices and edges must
be relabeled as the graph is reduced.
Doing this in random order leads to random access of
vertex and edge labels.
We try to optimize this by
either compressing the memory of the graph storage,
or by reordering the updates within each batch.
In the case that the tree is just a path,
much of the vertex and edge labeling can be done implicitly,
reducing the overhead.

\section{Heavy Path Graphs}
\label{sec:pathgraphs}
\begin{figure}[ht]
\centering
\includegraphics[scale=0.45]{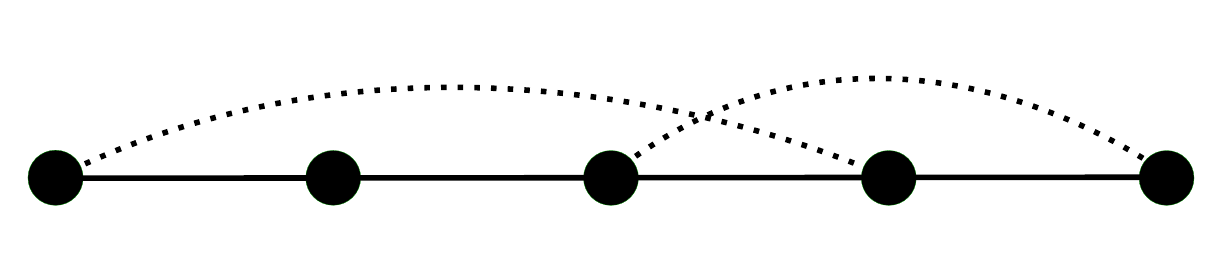}

\caption{An example of a heavy path graph.
The solid path edges are the low-stretch spanning tree
of the graph.}
\label{fig:heavypath}
\end{figure}

Here we introduce a class of model problems
that we will use to test and analyze different cycle-toggling approaches.
These graphs are constructed by adding edges between vertices 
on a path graph.
Edge resistances are selected so that the low-stretch spanning tree
of the resulting graph is always the underlying path.
As a consequence the edges on the path have larger edge weights
than the off-path edges, so we refer to this class of graphs
as \textit{heavy path graphs}.
An example of such a graph is shown in Figure~\ref{fig:heavypath}.

Our interest in these problems does not come from any
real world application.
Instead we believe these are natural models to consider
when studying KOSZ and other cycle-toggling algorithms.
We believe that this model can be tuned to have various stretch properties
along with spectral and graph separator properties, though we
do not explore that in this paper.
Furthermore they allow us to explore very fundamental questions
about data structures and cycle-toggling implementations.

This model simplifies many of the implementation issues
associated with dynamic trees, as the paths are easier
to handle than more general tree layouts.
Specifically, we can use a static, perfectly balanced
binary tree for the path.
This likely has the least data structure overhead
as the optimum separator of an interval is implicitly the
middle.
Furthermore, this allows us to store the tree in heap order,
which means the tree paths can be mapped to a subinterval
using bit operations, and the downward/upward propagations
can be performed iteratively.

\subsection{Example Models}
\label{subsec:examplemodel}
There are many possible subclasses that belong to the heavy
path graph model. We introduce several subclasses here
for experimentation.
\begin{enumerate}[(1),itemsep=-1ex]
\item \textbf{Fixed Cycle Length-1k}:
These graphs are composed of a tree path with random resistances between
1 and 10,000, combined with
off-tree edges between every pair $(i,i+1000)$, e.g.
an edge between vertices 1 and 1000, between vertices 2 and 1001, and so on.


\item \textbf{Fixed Cycle Length-2}:
These graphs are composed of a tree path with random resistances
between 1 and 10,000, combined with
off-tree edges between every pair $(i,i+2)$, e.g.
an edge between vertices 1 and 3, between vertices 2 and 4, and so on.


\item \textbf{Random Cycle Length}:
These graphs are composed of a tree path with random resistances
between 1 and 1000, combined with $n$ randomly selected off-tree edges,
where $n$ is the number of vertices.


\item \textbf{2D Mesh}:
These graphs embed a tree path in a 2D mesh.
The tree path resistances are chosen randomly between 1 and 1000.


\item \textbf{3D Mesh, Uniform Stretch}:
These graphs are similar to (4) but with a 3D mesh.


\end{enumerate}

We then consider two different ways of setting resistances on the
off-tree edges on all of the models above.
\begin{enumerate}[itemsep=-1ex]
\item \textbf{Uniform Stretch}
Resistances of off-tree edges are chosen so that stretch is 1 for every cycle.

\item \textbf{Exponential Stretch}
Resistances of off-tree edges are chosen so that cycles have stretch sampled from an
exponential distribution. 
\end{enumerate}

\section{Experiments}
\label{sec:experiments}

\subsection{Experimental Design}
We now describe empirical evaluations of the
cycle-toggling implementations
from Section~\ref{sec:algorithms} on the class of graphs
described in Section~\ref{sec:pathgraphs}.
As we only experiment on these path models, we can use
cycle-toggling methods that will only work on a path, but
we also employ their more general versions that will work
on any graph.
The four cycle-toggling implementations are as follows:
\begin{enumerate}[itemsep=-1ex]
\item BST-based data structure for general graphs
\item Path-only BST decomposition
\item Recursive divide-and-conquer for general graphs
\item Path-only recursive divide-and-conquer
\end{enumerate}
Additionally we implement a preconditioned conjugate
gradient with diagonal scaling to compare against
the cycle-toggling methods.
We implemented all of these in C++ and also have a Python/Cython
implementation of the general recursive method.
All algorithm implementations, graph generators, and test results
for this paper can be found at {\it https://github.com/sxu/cycleToggling}.
We also experimented with Hoske et al.'s~\cite{HoskeLMW15} implementation
of cycle-toggling.

We use all of the generators described in
Section~\ref{subsec:examplemodel} to create
different heavy path graphs with a varying total stretch.
We use vertex sizes of $5\times10^4$, $10^5$,
$5\times10^5$, and $10^6$. For the fixed cycle length generators, we set
$hop=1000$, and for the random cycle length generators, we set the
number of off-tree edges to $2n$. To get an idea for the various
stretch properties of these graphs, we list the total stretch
for size $10^6$ in Table~\ref{tab:stretch1e6}.
\begin{table}[htb!]
\centering
\small
\begin{tabular}{|l|c|c| }
\hline
 & Uniform & Exponential\\ \hline
Fixed Length-1k & 1.01e6 & 1.12e6\\\hline
Fixed Length-2 & 2.00e6 & 1.04e7 \\\hline
Random Length & 2.00e6 & 1.30e7\\ \hline
2D Mesh& 2.00e6 & 1.08e7\\\hline
3D Mesh & 3.82e6 & 2.27e7\\\hline
\end{tabular}
\caption{Total stretch for all graph models of size $10^6$.
For each of the model problems in~\ref{subsec:examplemodel}, this table shows
the total stretch of cycles formed by adding edges to the underlying path.
The models were generated with weights
to create cycles with uniform stretch (all cycles with stretch 1), and exponential
stretch(cycles with stretch chosen from an exponential distribution).}

\label{tab:stretch1e6}
\end{table}

We also generate right hand side vectors
$\bb$ in two different ways to obtain both local and global behaviors.
\begin{enumerate}[itemsep=-1ex]
\item Random: Randomly select $\xx$ and form $\bb = \LL \xx$,
\item (-1,1): Pick $\bb$ to route $1$ unit of electrical flow from
the left endpoint of the path to the right endpoint.
\end{enumerate}

Experiments were performed on Mirasol, a shared memory
machine at Georgia Tech, with 80 Intel(R) Xeon(R) E7-8870 processors
at 2.40GHz. Problems were solved to a residual tolerance of
$10^{-5}$.

\subsection{Experimental Results}
\begin{figure}[ht]
\centering
\includegraphics[scale=\figsize]{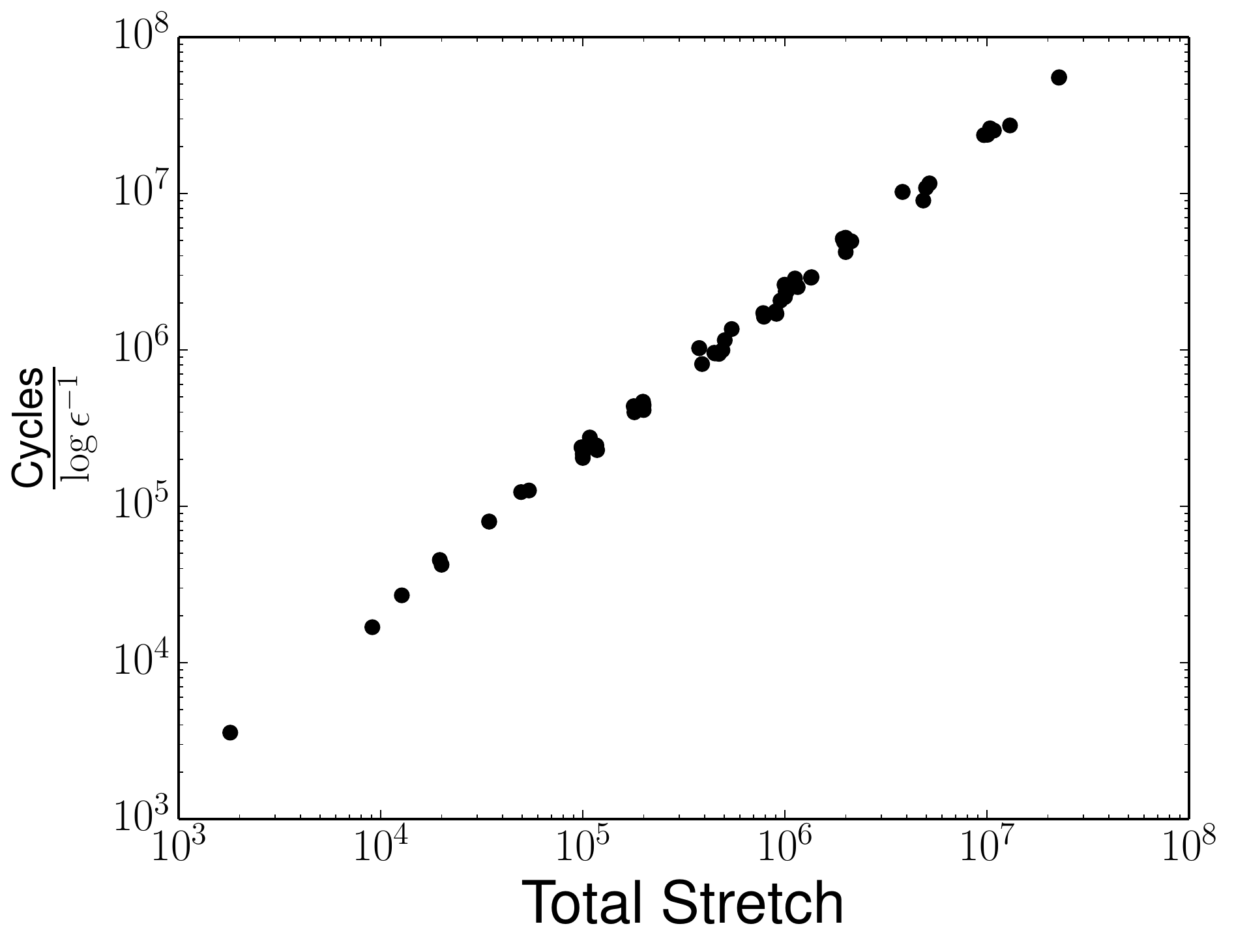}
\caption{KOSZ asymptotic dependence on tree stretch. The number
of toggles required by KOSZ is shown as a function
of tree stretch. The reasonable slope indicates a lack of large
hidden constants in KOSZ complexity.}
\label{fig:stretch}
\end{figure}

We first examine the asymptotic behavior of the cycle-toggling
methods on all the test graphs. Figure~\ref{fig:stretch}
shows the number of cycles required for convergence as a function
of total stretch. This figure only involves solves using the 0-1
right hand side as this was always a more difficult case.

We omit results from the Hoske et al.\ implementation because we found its
performance to be slower by a factor of $50$ than our cycle-toggling implementations.
Their initialization costs are much higher than solve costs, making it prohibitively
expensive to run on all of the test graphs in our set.

To visualize the comparison of cycle-toggling implementations
on all the different test graphs, we utilize a performance profile plot shown in
Figure~\ref{fig:perfprofile}.
A performance profile~\cite{DolanM02} calculates, for some performance metric,
the relative performance
ratio between each solver and the best solver on every problem instance.
In our case the metric of interest is the average cycle-toggle time, so for each
method and every graph, the relative performance ratio is the method's average
cycle-toggle time divided by the lowest average cycle-toggle time over all methods.
Then to capture how a method fares across the entire problem set, the
performance profile
shows the fraction of test problems (on the y-axis) that are within a distance
(on the x-axis) from the relative performance ratio.
This plot contains all
the different model problems at every problem size tested.

\begin{figure}[ht]
\centering
\includegraphics[scale=\figsize]{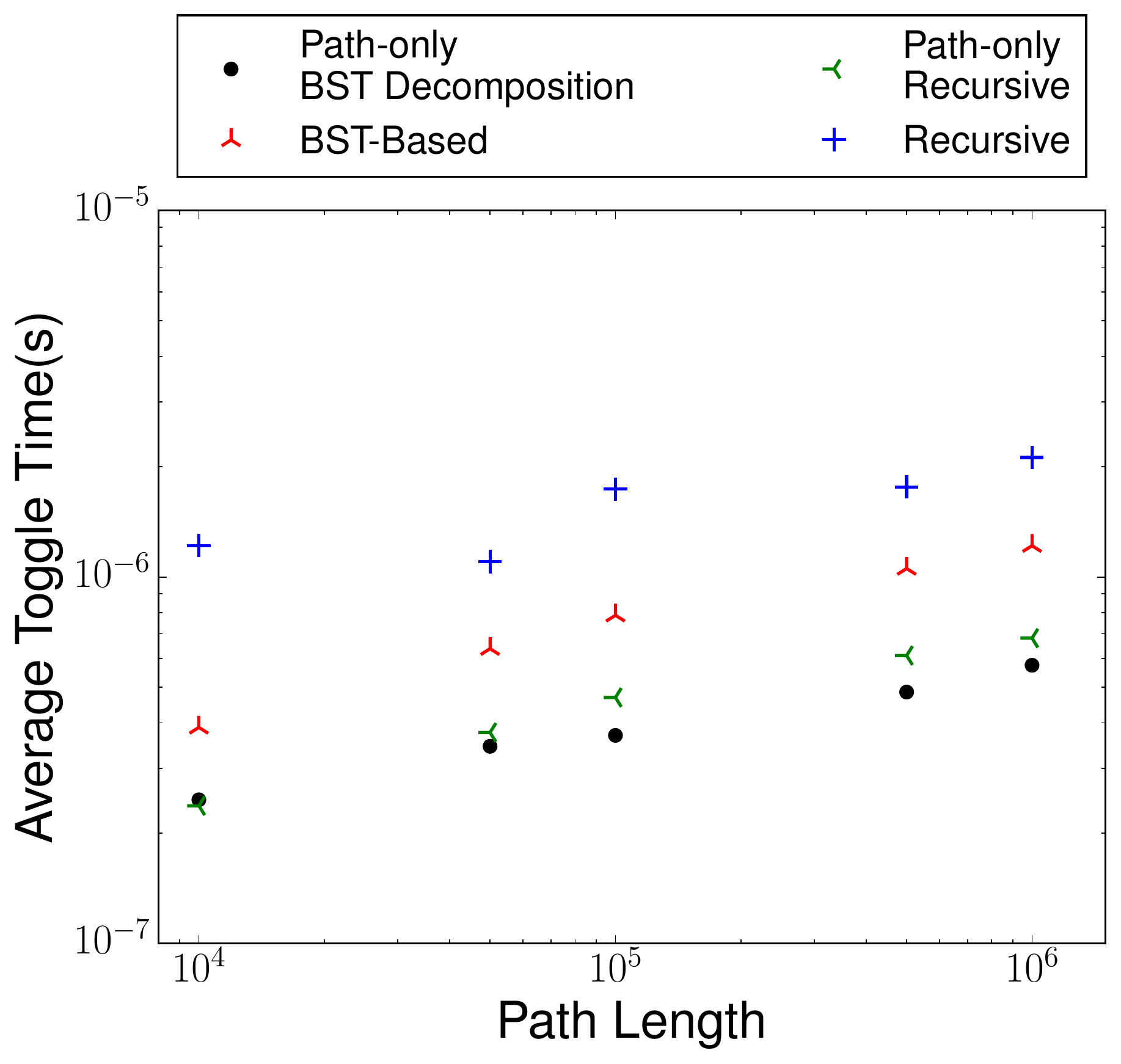}
\caption{Weak scaling of cycle-toggle performance of all methods on unweighted
3D mesh model problems with exponential stretch. Average cycle-toggle time is
shown as a function of problem size where an upward slope indicates decreased
performance with larger problem size.}
\label{fig:weakscaling}
\end{figure}

Weak scaling experiments, measuring cycle-toggle
performance as graph size increases, are useful
for predicting performance on larger problems.
The scaling behavior was relatively similar across
the model problems so we only show one example
in Figure~\ref{fig:weakscaling} for
the 3D Unweighted Mesh with exponential stretch.

\begin{figure}[ht]
\centering
\includegraphics[scale=\figsize]{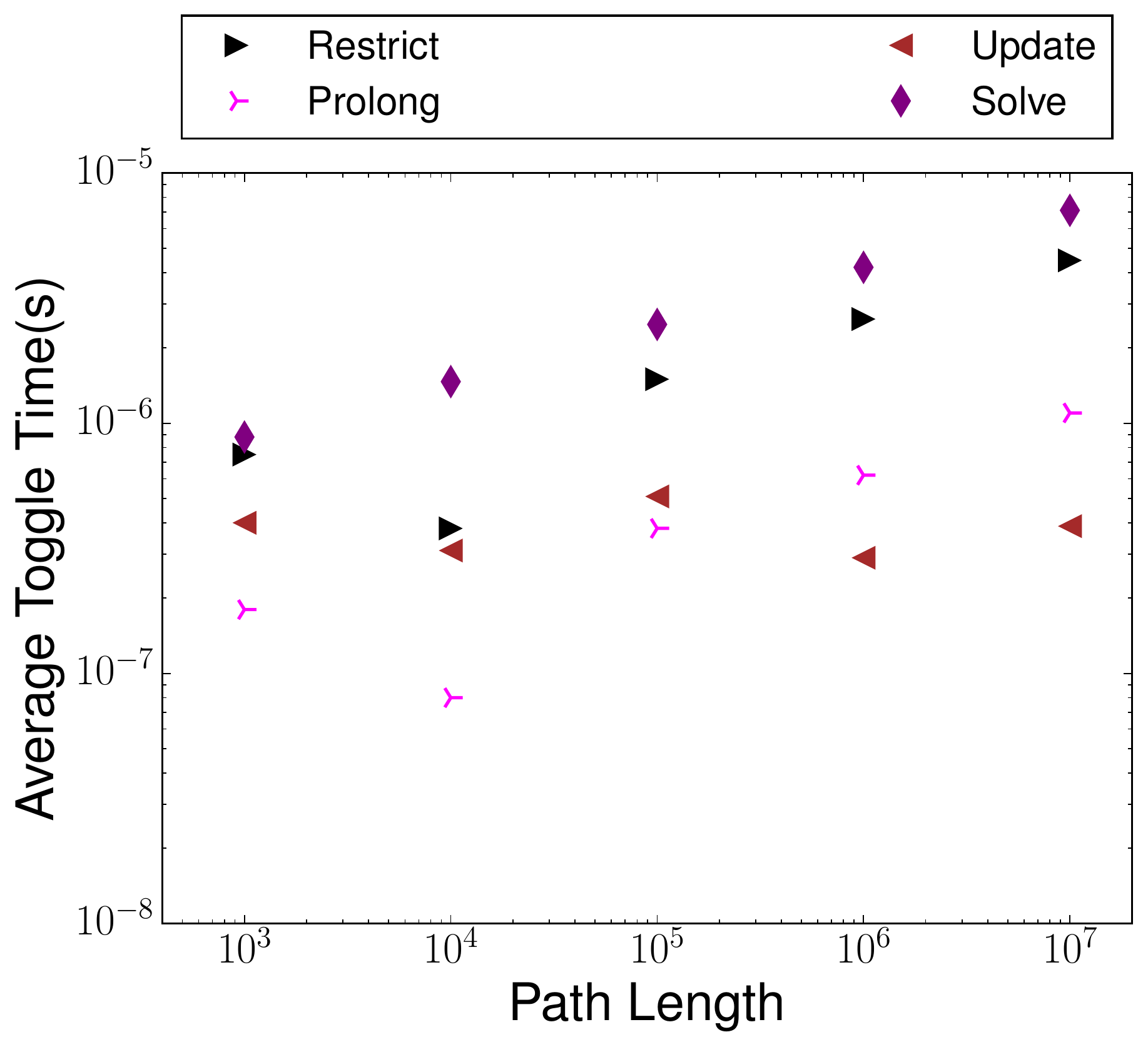}
\caption{Weak scaling of cycle-toggle performance for the recursive solver
on FixedLength-1k model problems.
Average cycle-toggle time is shown along with its most expensive
sub-components: restriction, solve, and prolongation. Upward slopes
indicate decreasing performance with problem size. 
}
\label{fig:hierarchytime}
\end{figure}

We examine how much time the recursive method spends restricting
and prolonging flow in the recursive hierarchy, and how much time
is spent doing cycle-toggles in Figure~\ref{fig:hierarchytime}.
Results are shown for the FixedLength-1k model with a slightly wider
range of problem size than the other experiments.
The solve time in this plot includes the sum of the other operation timings,
along with memory allocation.
We did this profiling with our Python/Cython implementation, but we believe
the C++ performance is comparable. 

\begin{figure}[htb!]
\centering

\includegraphics[scale=\figsize]{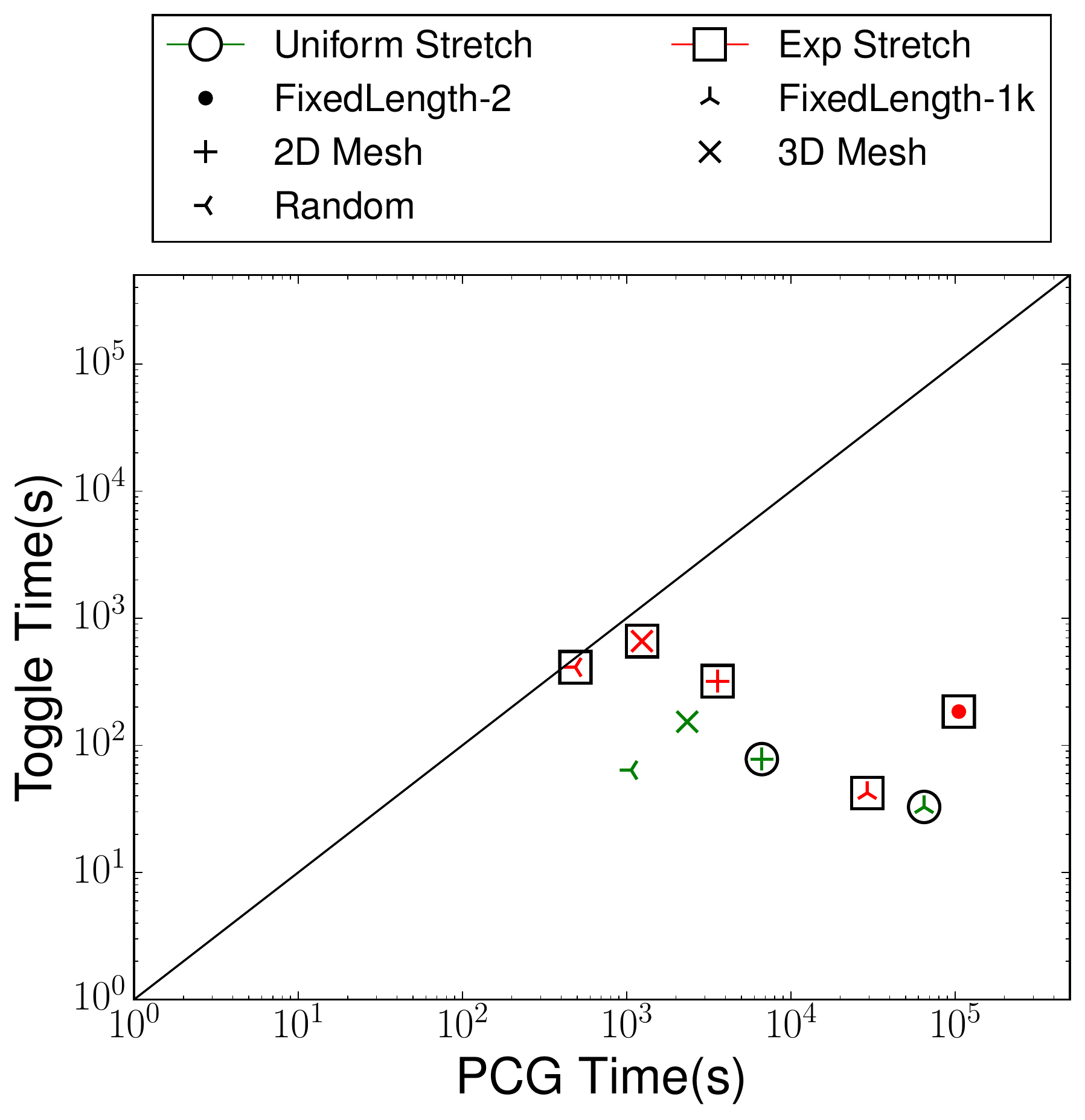}
\caption{Comparison of BST-based data structure cycle-toggling to PCG
by graph type.
Points under the line indicate cycle-toggling method outperformed
PCG.}
\label{fig:pcgcompare}
\end{figure}

Figure~\ref{fig:pcgcompare} shows BST-based
cycle-toggle timing results
relative to PCG results.
Points below the line indicate cycle-toggling was faster,
while points above the line are slower.
This plot only includes size $10^6$ problems using the 0-1 right
hand side. A random right hand side plot is omitted for space
as these problems were much easier for both solvers, though slightly
relatively easier for PCG.




\subsection{Experimental Analysis}
In Figure~\ref{fig:stretch} the cycle-toggling
methods' asymptotic dependence on tree stretch
is near constant with a slope close to 1.
Note that this plot
would be linear even without the log axes.
Concerning KOSZ practicality, it is highly important to see that there is not a
large slope, which would indicate a large hidden constant in the
KOSZ cost complexity. This plot tells us that with
a combination of low-stretch trees and fast cycle update methods,
dual space algorithms have potential.
This figure also helps illustrate the
range of problems we are using for these experiments.
The stretch and resulting cycle cost both vary between
four to five orders of magnitude.

The performance profile in Figure~\ref{fig:perfprofile} indicates that the data
structure based cycle-toggling methods performed the best using our
implementations. For the path-only BST decomposition,
the fraction of problems is already at 1 for a relative performance distance of 1,
meaning that this was always the fastest.
The path-only recursive method was slower, but still typically performed
better than the general implementations, being half as fast as
the path-only BST  method on 60\% of the problems.
Comparing the two general implementations, the tree data structure is within
a factor 4 of the best on 80\% of the problems, whereas the recursive method
is only within a factor of 4 on 40\% of the problems.  
A distance of 10 indicates performance within the same
order of magnitude, which the general
recursive method achieved on 80\% of the problems, indicating
that these methods are competitive with one another.

The weak scaling experiments shown in Figure~\ref{fig:weakscaling}
do indicate a decrease in cycle-toggle performance as graph size
increases. However, this plot is fairly optimistic, the largest
performance decrease is about $2.5 \times$ as the graph size
increases two orders of magnitude. The non steady plot
for the general recursive solver probably indicates that
the batch sizes were not scaled appropriately. Again, this plot
is only for one of the graph models, but most of them looked very
similar to this.

Figure~\ref{fig:hierarchytime} helps identify the performance
bottlenecks of the recursive method.
The actual time spent updating cycles is less than
the restriction and prolongation time.
The restriction time is by far the most expensive,
as it also includes time for relabeling edges and vertices.
The scaling of this plot shows a stable update cost,
with increasing restriction and prolongation costs.
This method was designed to keep the update costs
stable while increasing problem size, which seems
to be case. Unfortunately the restriction and prolongation
overhead costs are large and growing with problem size.
Still, these operations are not highly optimized, and we
wonder if we can borrow techniques from the multigrid community
to speed them up.

The PCG experiments in Figure~\ref{fig:pcgcompare} indicate
that cycle-toggling can outperform PCG on these heavy path models,
using the 0-1 right hand side.
This class of problems had a wider performance
gap for PCG than for the cycle-toggling routines, by about
an order of magnitude. Furthermore, the graph property that causes
difficulty for the solvers is different in each case;
cycle-toggling has trouble on the graphs with exponential stretch,
while PCG has difficulty with the fixed cycle length problems
(FixedLength-2 with uniform stretch even failed).
These results suggest that heavy path graphs are a good
direction to explore while searching for problems which
could benefit from cycle-toggling methods.

\section{Discussion and Conclusion}
\label{sec:conclusion}

We studied two approaches for implementing cycle-toggling based
solvers, data structures and recursive divide-and-conquer.
Using the heavy path model, we experimented on problems that are
are conceptually simple, but provide a range of solve behavior
through varying graph structure and stretch.
The recursive cycle-toggling was not as fast as the data structure
approach, but was still competitive, being in the same order
of magnitude on most problems.
method to general graphs, exhibited competitive behaviors.
Also both methods scaled reasonably with problem size.

While these experiments are a good start, there are several
directions we hope to continue this work.
The recursive update approach is outperformed
by the BST-based data structure approach in timing experiments.
We hope to complement these
results with floating point operation measurements.
We don't claim to have optimized the graph contraction,
flow restriction/prolongation, or cycle updates. Measuring the
number of operations the recursive solver spends on these would
help indicate fundamental performance.

The heavy path graphs
are a great model problem for seeing the effect path
resistances have on solver behavior.
They also allow us set aside the issue of finding
a low stretch spanning tree to focus instead
on the cost per cycle update.
We plan to continue modifying
these path resistances and initial vertex demands to find
interesting test cases.
However, for these methods
to be useful in practice we must extend them to more general classes
of graphs.

Dual cycle-toggling Laplacian solvers have until now been considered mainly
in the realm of theory.
Our comparisons of these methods to PCG indicate that there
are problems for which the dual methods can be useful.
In the future, we plan to combine primal and dual methods,
trying to get the best of both worlds.



\begin{thebibliography}{99}

\bibitem{Axelsson94:book}
O. Axelsson, {\em Iterative solution methods}, Cambridge University Press, New York, NY, 1994.

\bibitem{BenderDF00}
M.~A. Bender, E.~D. Demaine, and M. Farach-Colton, {\em Cache-oblivious B-trees}, IEEE FOCS, Redondo Beach, CA, 2000, pp.~399--409.

\bibitem{BomanDG15}
E.~G. Boman, K. Deweese, and J.~R. Gilbert, {\em Evaluating the dual randomized Kaczmarz Laplacian linear solver}, Informatica, 40(1) (2016), pp.~95--107.

\bibitem{BomanDG16}
E.~G. Boman, K. Deweese, and J.~R. Gilbert, {\em An empirical comparison of graph Laplacian solvers}, SIAM ALENEX, Arlington, VA, 2016, pp.~174--188.

\bibitem{BomanHV04}
E.~G. Boman, B. Hendrickson, and S. Vavasis, {\em Solving elliptic finite element systems in near-linear time with support preconditioners}, SIAM J. on Numerical Anal., 46(6) (2008), pp.~3264--3284.

\bibitem{BriggsM00}
W.~L. Briggs, V.~E. Henson, and S.~F. McCormick, 
{\em A multigrid tutorial}, SIAM, 2000.


\bibitem{CoFMNPW14} 
M.~B. Cohen, B.~T. Fasy, G.~L. Miller, A. Nayyeri, R. Peng, and N. Walkington, {\em Solving $1$-{L}aplacians of convex simplicial complexes in nearly linear time: collapsing and expanding a topological ball}, SIAM SODA, Portland, OR, 2014, pp.~204--216.

\bibitem{ChristianoKMST11}
P. Christiano, J.~A. Kelner, A. Madry, D.~A. Spielman, and S.-~H. Teng, {\em Electrical flows, Laplacian systems, and faster approximation of maximum flow in undirected graphs}, ACM STOC, San Jose, CA, 2011, pp.~273--282.

\bibitem{CohenKMPPRX14}
M.~B. Cohen, R. Kyng, G.~L. Miller, J.~W. Pachocki, R. Peng, A. Rao, and S.~C. Xu, {\em Solving {SDD} linear systems in nearly {\it m}log$^{{1/2}}${\it n} time}, ACM STOC, San Jose, CA, 2011, pp.~343--352.

\bibitem{CormenLRS09}
T.~H. Cormen, C.~E. Leiserson, R.~L. Rivest, and C. Stein, {\em Introduction to algorithms}, MIT Press and McGraw-Hill, 2009.

\bibitem{ChinMMP13}
H.~H. Chen, A. M{{a}}dry, G.~L. Miller, and R. Peng, {\em Runtime guarantees for regression problems}, ITCS, Berkeley, CA, 2013, pp.~269--282.

\bibitem{DolanM02}
E.~D. Dolan and J.~J. Mor{\'e}, {\em Benchmarking optimization software with performance profiles}, Mathematical Programming, 91(2) (2002), pp.~201--213.

\bibitem{DoyleS84:book}
P.~G. Doyle and J.~L. Snell, {\em Random walks and electric networks}, Mathematical Association of America, 1984.

\bibitem{HoskeLMW15} D. Hoske, D. Lukarski, H. Meyerhenke, and M. Wegner, {\em Is nearly-linear time the same in theory and practice? A case study with a combinatorial Laplacian solver}, SEA, Paris, FRA, 2015, pp.~205--218.

\bibitem{KyngLPSS15:arxiv}
R. Kyng, Y.~T. Lee, R. Peng, S. Sachdeva, and D.~A. Spielman, {\em Sparsified Cholesky and multigrid solvers for connection Laplacians}, Computing Research Repository, 2015, http://arxiv.org/abs/1512.01892.


\bibitem{KoutisMP10:journal}
I. Koutis, G.~L. Miller, and R. Peng, {\em Approaching optimality for solving {SDD} systems}, SIAM J. on Comp., 43(3) (2014), pp.~337--354.

\bibitem{KoutisMP11}
I. Koutis, G.~L. Miller, and R. Peng {\em A Nearly-m log n time solver for {SDD} linear systems}, IEEE FOCS, Palm Springs, CA, 2011, pp.~590--598.

\bibitem{KelnerOSZ13}
J.~A. Kelner, L. Orecchia, A. Sidford, and Z.~A. Zhu, {\em A simple, combinatorial algorithm for solving SDD systems in nearly-linear time}, ACM STOC, Palo Alto, CA, 2013, pp.~911--920.

\bibitem{KyngRS15}
R. Kyng, A. Rao, and S. Sachdeva, {\em Fast, provable algorithms for isotonic regression in all $\ell_p$-norms}, NIPS, Montreal, QC, 2015, pp.~2701--2709.

\bibitem{LeeRS13}
Y.~T. Lee, S. Rao, and N. Srivastava, {\em A new approach to computing maximum flows using electrical flows}, ACM STOC, Palo Alta, CA, 2013, pp.~755--764.

\bibitem{LeeS13}
Y.~T. Lee and A. Sidford, {\em Efficient accelerated coordinate descent methods and faster algorithms for solving linear systems}, IEEE FOCS, Berkeley, CA, 2013, pp.~147--156.

\bibitem{LeeS14}
Y.~T. Lee and A. Sidford, {\em Path finding methods for linear programming: solving linear programs in $\tilde{O}(\sqrt{rank})$ iterations and faster algorithms for maximum Flow}, IEEE FOCS, Philadelphia, PA, USA, 2014, pp.~424--433. 

\bibitem{Madry13}
A. Madry, {\em Navigating central path with electrical flows: from flows to matchings, and back}, IEEE FOCS, Berkeley, CA, 2013. pp~253--262.

\bibitem{PengS14}
R. Peng and D.~A. Spielman, {\em An efficient parallel solver for {SDD} linear systems}, ACM STOC, New York, NY, USA, 2014, pp.~333--342.

\bibitem{ReMiMo93}
M. Reid-Miller, G.~L. Miller, and F. Modugno, {\em List ranking and parallel tree contraction} in J.~H. Reif {\em Synthesis of parallel algorithms}, Morgan Kaufmann, San Francisco, CA, 1993, pp.~115--194.

\bibitem{Saad03}
Y. Saad, {\em Iterative methods for sparse linear systems}, SIAM, 2003.

\bibitem{SleatorT83}
D.~D. Sleator and R.~E. Tarjan, {\em A data structure for dynamic trees}, J. Comp. Syst. Sci., 26(3) (1983), pp.~362--391.

\bibitem{SpielmanTengSolver:journal}
D.~A. Spielman and N. Srivastava, {\em Graph sparsification by effective resistances},
SIAM J. on Comp., 40(6) 2011, pp.~1913--1926.

\bibitem{SpielmanTengSolver:journal}
D.~A. Spielman and S.-~H. Teng, {Nearly linear time algorithms for preconditioning and solving symmetric, diagonally dominant linear systems}, SIAM J. on Matrix Anal. and Appl.,
35(3) 2014, pp.~835--885.

\bibitem{TolliverMiller06}
D. Tolliver and G.~L. Miller, {\em Graph Partitioning by Spectral Rounding: Applications in Image Segmentation and Clustering}, IEEE CVPR, New York, NY, 2006, pp.~1053--1060.

\end{thebibliography}


\end{document}